\documentclass[sigconf,nonacm]{acmart}

\usepackage{array}
\usepackage{booktabs}
\usepackage{enumitem}
\usepackage{graphicx}
\usepackage{microtype}
\usepackage{multirow}
\usepackage{pifont}
\usepackage{stfloats}
\usepackage{tabularx}

\setcopyright{none}
\acmDOI{}
\acmISBN{}
\setlist[itemize]{leftmargin=*,nosep}

\newcolumntype{Y}{>{\raggedright\arraybackslash}X}
\newcommand{\ScreenYes}{\ding{51}}
\newcommand{\ScreenNo}{\textemdash}

\newcommand{\EvalIntervalCount}{9}
\newcommand{\EvalStateRows}{2686}

\newcommand{\FallRecognitionAccuracyPct}{96.5}
\newcommand{\AgentEventPassed}{58}
\newcommand{\AgentEventCases}{60}
\newcommand{\AgentFusionCaseCount}{24}
\newcommand{\AgentFusionPassed}{23}
\newcommand{\AgentDetPNinetyFiveMs}{0.56}
\newcommand{\AgentChecklistItems}{213}
\newcommand{\AgentChecklistPassed}{206}
\newcommand{\AgentChecklistPassPct}{96.7}
\newcommand{\AgentDiagnosisScenarioCount}{24}
\newcommand{\AgentDiagnosisPassed}{22}
\newcommand{\AgentTracePassed}{23}
\newcommand{\AgentDiagnosisModelName}{GPT-5.4}
\newcommand{\AgentDiagnosisPNinetyFiveMs}{6560.7}

\begin{document}

\title{\textbf{SuperSenseDoctor}: A Multimodal and Contactless Agent for Health Tracking}

\author{Xuwen Zhang, Zijian Lu, Yicheng Lei, Rui Qiu, Jiale Li, Yiping Zuo, Weibei Fan, and Fu Xiao}
\authornote{Yiping Zuo, Weibei Fan, and Fu Xiao are faculty advisors.}
\affiliation{%
  \institution{Nanjing University of Posts and Telecommunications}
  \city{Nanjing}
  \country{China}
}
\renewcommand{\shortauthors}{Zhang et al.}

\begin{abstract}
Population aging is increasing the need to monitor older adults safely and independently at home. However, cameras, wearables, and manual checks often introduce privacy, adherence, and attention burdens that hinder sustained health monitoring. This paper presents \textbf{SuperSenseDoctor}, a multimodal contactless agent architecture for long-term home health tracking. The system transforms WiFi, mmWave radar, and surface temperature into a persistent human health state. The system relies on fixed decision rules to conduct continuous daily monitoring and respond to pre-defined hazards. When abnormal signals appear, event-driven reasoning analyzes only standardized evidence to produce traceable care-support measures. In this manner, \textbf{SuperSenseDoctor} integrates sensing, temporal state, reasoning, and action into a unified and auditable loop. The calibrated multimodal pipeline achieves 1.994~bpm mean absolute error (MAE) and 3.142~bpm root mean square deviation (RMSD) for heart rate, 0.197~bpm MAE and 0.263~bpm RMSD for respiratory rate, and \FallRecognitionAccuracyPct\% fall-recognition accuracy. The evaluation also covers \EvalStateRows\ one-second states across \EvalIntervalCount\ chronological intervals and reaches a \AgentChecklistPassPct\% criterion-level Agent checklist pass rate. These results demonstrate the feasibility of a stateful contactless sensing-to-action architecture for long-term home health monitoring.
\end{abstract}


\maketitle
\raggedbottom
\begin{figure*}[t]
	\centering
	\includegraphics[width=0.98\textwidth]{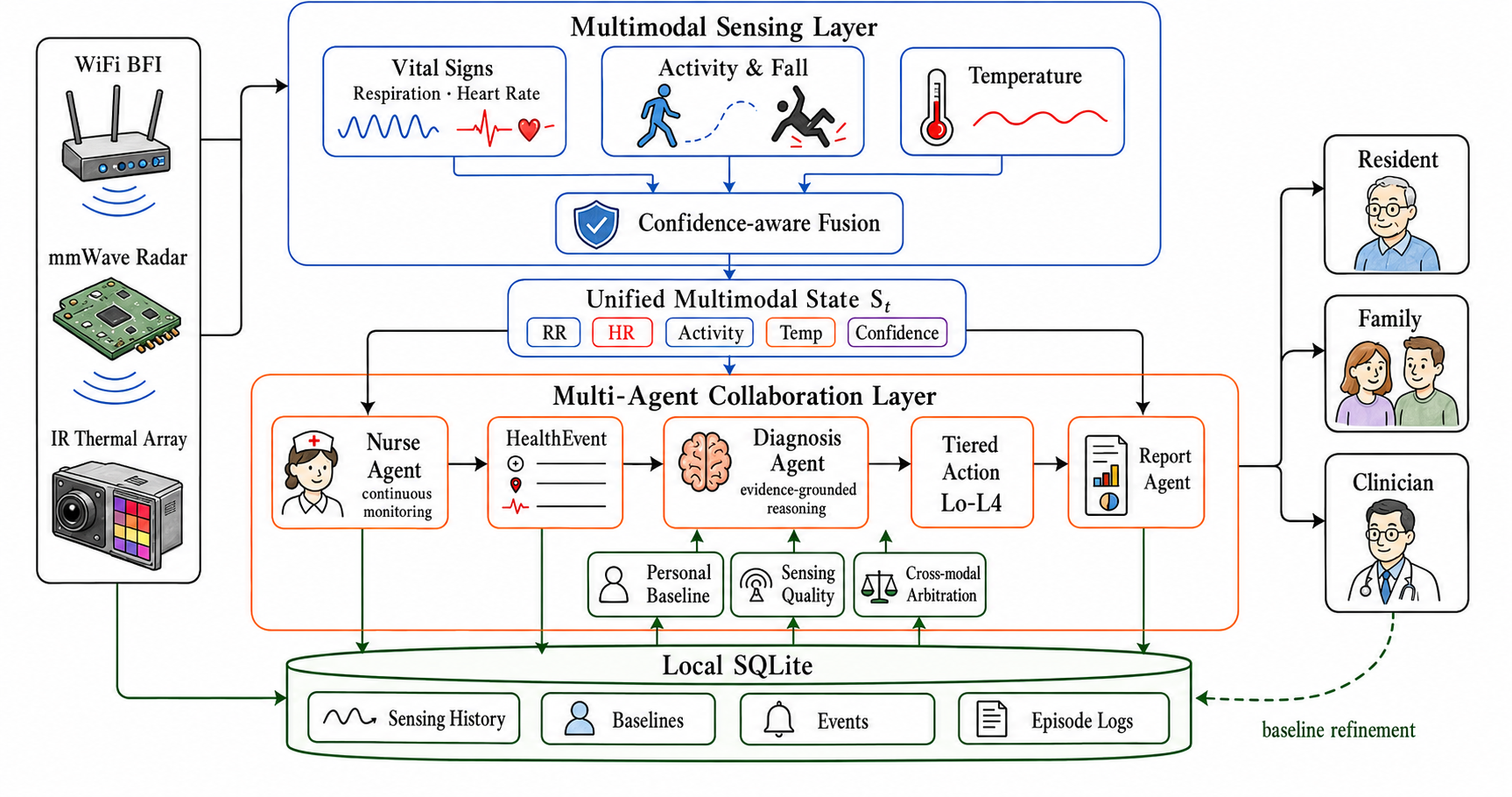}
	\caption{SuperSenseDoctor system design. Contactless modalities become a confidence-aware state that triggers deterministic monitoring, evidence-grounded Agent reasoning, tiered action, and reporting.}
	\label{fig:architecture}
	\Description{A hand-drawn architecture diagram connecting WiFi BFI, mmWave radar, and an infrared thermal array to multimodal sensing, a unified state, Nurse, Diagnosis, and Report Agents, tiered actions, local SQLite storage, and resident, family, and clinician outputs.}
\end{figure*}

\section{Introduction}

Imagine an older adult living alone begins to deviate from their usual routine. Behavioral sensing, thermal trends, WiFi Beamforming Feedback Information (WiFi BFI), and mmWave radar each provide partial evidence, but no single stream is conclusive. Home health tracking therefore needs contactless measurements that remain separate enough to explain uncertainty and connected enough to support timely care action.

The sensing pieces for this requirement are now available. Reflected radio signals support vital monitoring and device-free activity recognition \cite{adib2015vitalradio,wang2020phasebeat,wang2015carm,zheng2019widar3}, while millimeter-wave systems add focused vital-sign estimation and fall detection without cameras \cite{yang2016mmvital,jin2022mmfall}. Digital-health studies also show that respiratory rhythm, heart dynamics, gait, routine, and temperature can act as screening cues across several chronic and acute conditions \cite{chen2025copd,kapur2017osa,sequeira2019osahrv,boehmer2017multisense,sun2024digitalpd,popp2024passivead,cepukaityte2024dementia,mason2022tempredict}. Yet sensing availability alone does not solve the system problem because conventional fusion can hide cross-modal evidence behind one estimate, while sending raw streams to a language model is costly and difficult to audit.

The missing piece is not another isolated sensor classifier. Multimodal learning and trusted multi-view inference study how to combine complementary observations under uneven quality \cite{baltrusaitis2019multimodal,wang2020gradientblending,han2021trustedmultiview}, while recent Agent systems explore tool use, reasoning, and medical collaboration \cite{yao2023react,liu2024agentbench,kim2024mdagents,fan2025aihospital}. Our prototype observations show why this connection must remain inspectable. WiFi BFI and mmWave do not always agree on heart rate or respiration when posture, bedding, or line-of-sight (LOS) conditions change. \textbf{SuperSenseDoctor} therefore turns contactless room sensing into a persistent health state that keeps values, confidence, missingness, non-line-of-sight (NLOS) state, and event provenance available before an Agent proposes care-support actions.

\textbf{SuperSenseDoctor} addresses this gap through four connected stages of multimodal sensing, a structured state object, event-driven Agent collaboration, and an auditable action proposal. WiFi BFI provides wide-area motion and periodic-signal cues, mmWave adds localized chest-motion and fall evidence, and a thermal-array interface supplies body-surface temperature. The system preserves per-modality evidence and quality events so downstream components can distinguish possible health changes from unreliable sensing.

Figure~\ref{fig:architecture} shows this data path. The core insight is that an Agent is useful here as an evidence gate between sensing uncertainty and care action. Normal operation stays deterministic, while richer reasoning is reserved for rules or quality events that create a question. Each output keeps the event, queried evidence, uncertainty, proposed tier, and resulting channel available for inspection. Our contributions are listed below.
\begin{itemize}
    \item \textbf{Multimodal Sensing}. We fuse WiFi BFI, mmWave radar, and thermal sensing into a contactless health state covering respiratory rhythm, cardiac motion, falls, activity, and temperature.
    \item \textbf{Agent Layer}. We build Nurse, Diagnosis, and Report Agents that turn standardized sensing evidence into traceable care actions with temporal provenance.
    \item \textbf{Experimental Results}. The prototype reaches heart-rate MAE and RMSD of 1.994 and 3.142~bpm, respiratory-rate MAE and RMSD of 0.197 and 0.263~bpm, \FallRecognitionAccuracyPct\% fall-recognition accuracy, complete state coverage across \EvalIntervalCount\ chronological intervals, and a \AgentChecklistPassPct\% criterion-level Agent checklist pass rate.
\end{itemize}


\vskip -0.7em
\section{\textbf{SuperSenseDoctor}}

\subsection{Multimodal State Fusion}

\textbf{SuperSenseDoctor} treats multimodal sensing as an arbitration problem rather than a simple averaging problem. WiFi BFI, mmWave radar, thermal sensing, and behavioral sensing observe the same resident through different physical channels. Their errors are not identical. Bedding may weaken radar, body motion may disturb WiFi, and thermal trends may lag vital-sign changes. The core goal is to estimate vital signs while retaining enough disagreement evidence for the Agent to reason about sensing quality.

WiFi BFI and mmWave radar provide complementary radio evidence. WiFi BFI works with commodity WiFi hardware, covers a wider room area, and is more practical for engineering deployment than CSI. It is therefore useful for long-term, low-intrusion monitoring and activity disturbance cues. mmWave radar provides a more localized view of chest micro-motion and body movement, which supports finer respiratory, cardiac, and fall-related evidence. Their combination gives the state both broad contextual coverage and local motion detail.

Each adapter first converts raw observations into candidate evidence. For vital signs, mmWave and WiFi BFI apply PSR-DFE subcarrier selection and IVY-SVMD decomposition to isolate respiratory and cardiac periodic components, then generate heart-rate and respiratory-rate candidates under contactless periodic-motion principles \cite{adib2015vitalradio,yang2016mmvital}. For falls, the system checks short-window velocity, height, and posture for abrupt changes, with WiFi BFI supplying motion cues and mmWave supplying localized body-motion evidence. Thermal and behavioral streams provide temperature, activity, posture, and routine context. For each second $t$ and modality $m$, the adapter emits an evidence token $e_t^m$ with four fields: candidate value $v_t^m$, confidence $c_t^m$, quality $q_t^m$ for missingness, NLOS, or conflict, and provenance $p_t^m$.

The \texttt{FusionEngine} then performs confidence-aware arbitration for each metric. It first builds a reliable-modality set from confidence and NLOS constraints. If reliable radio branches agree within tolerance, the result is a confidence-weighted mean. If only one branch remains reliable, that branch dominates. If two plausible branches conflict, \textbf{SuperSenseDoctor} does not hide the conflict behind one estimate. It emits a quality event and preserves the branch-level evidence. This event-preserving fusion is the main difference from a conventional early-fusion pipeline: uncertainty becomes a first-class signal for downstream reasoning.

The temporal updater stores the selected values and unresolved evidence in the unified multimodal state $S_t$. In implementation, $S_t$ is the \texttt{StateObject}. It carries respiratory rate, heart rate, activity, temperature, confidence, posture, fall status, quality events, and provenance. Thermal and behavioral evidence stay as context rather than being collapsed into the vital-sign estimate. The Agent reads a state with arbitration history, not one opaque number.

\vskip -0.5em
\subsection{Event-Driven Agent Layer}

The Agent layer turns state estimation into care-support reasoning through event-gated computation.

The Nurse Agent is a deterministic first screen over $S_t$. It compares each state with project thresholds, personal baselines, persistence, and sensing-quality flags. Normal windows update local history without invoking the LLM. Falls, baseline shifts, extreme respiratory values, modality conflicts, weak dual sensing, and NLOS degradation become typed \texttt{HealthEvent}s. High-risk combinations take a reflex path. Other events enter the Diagnosis Agent with a bounded evidence package.

The Diagnosis Agent is designed as a virtual triage doctor over a bounded evidence graph. It does not receive raw streams. For each \texttt{HealthEvent}, registered tools retrieve personal baseline, sensing quality, cross-modal arbitration, recent events, and the latest state. A compact project-policy card defines the L0--L4 action mapping. The Agent must ground its answer in these tool returns and produce a typed \texttt{TriageDecision}. The decision records the action tier, interpretation, evidence anchors, uncertainty, and channel. L0 and L1 mean routine observation. L2 prompts the resident. L3 notifies family. L4 routes to an emergency channel with human confirmation. This design makes the LLM a bounded reviewer of standardized evidence rather than an open diagnostic oracle.

The Report Agent closes the loop by converting isolated decisions into longitudinal memory. Each \texttt{EpisodeLog} links the trigger to the sensing summary, triage tier, uncertainty, action channel, tool use, and reflex status. SQLite keeps sensing history, baselines, events, and episode logs together. Resident prompts, family notifications, and clinician review therefore share the same evidence trail. Clinician review can refine personal baselines for later monitoring, which feeds back into the Nurse Agent's future event thresholds.

\section{Implementation}

\begin{figure}[t]
\centering
\includegraphics[width=\columnwidth]{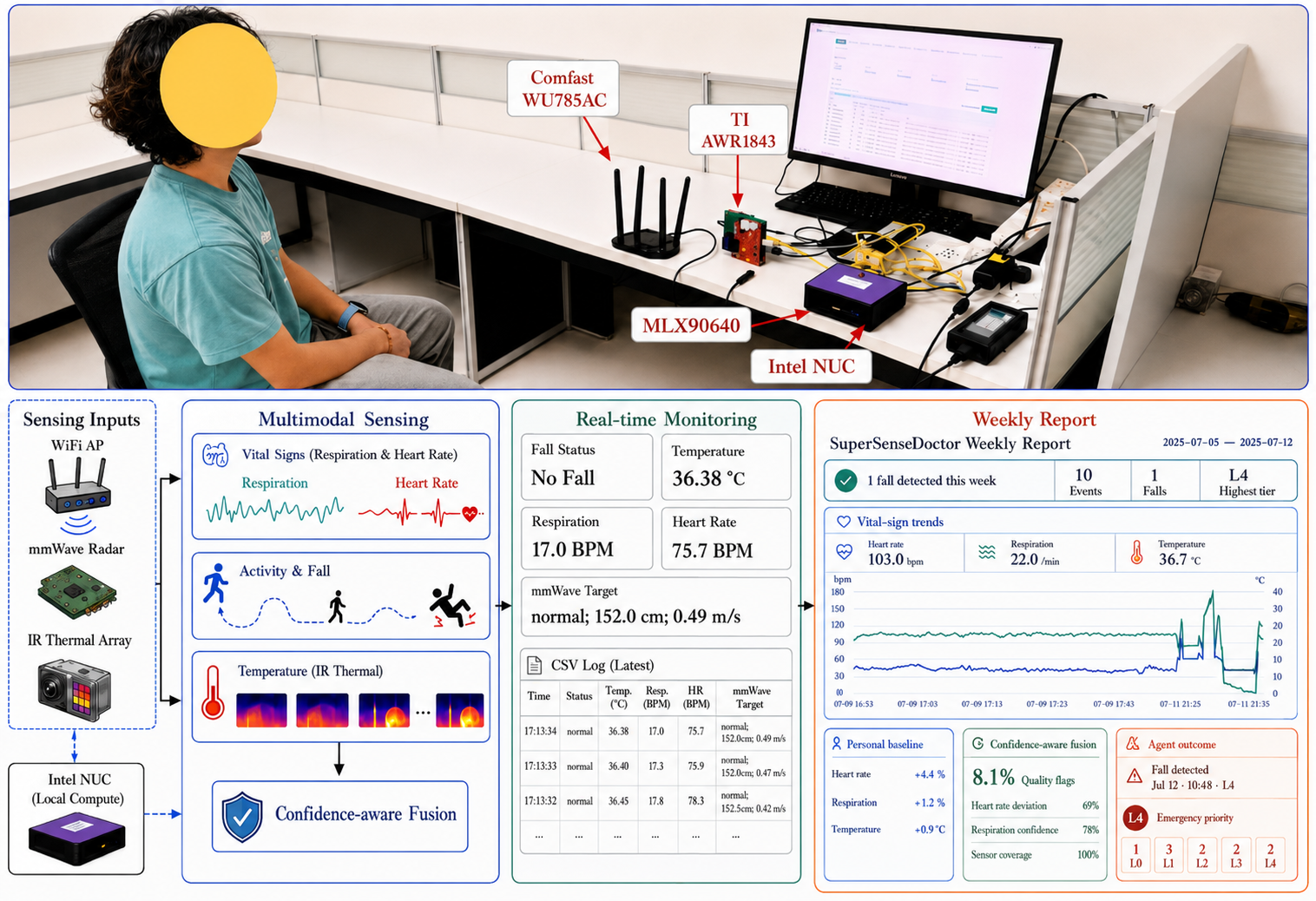}
\caption{Prototype deployment and sensing-to-report workflow.}
\Description{Composite prototype figure showing a seated participant near the WiFi AP, mmWave radar, thermal sensor, and local compute host, followed by panels for sensing inputs, multimodal sensing, real-time monitoring, and a weekly report dashboard.}
\label{fig:device}
\end{figure}

Figure~\ref{fig:device} shows the room-side prototype and the sensing-to-report workflow. Comfast WU785AC supplies WiFi BFI evidence for periodic motion and fall-related disturbance cues. TI AWR1843 adds radar evidence. MLX90640 supplies thermal input. The Intel NUC hosts the local hub and Agent runtime.

The software prototype is implemented in Python as an asynchronous event chain. A sensor aligner constructs state objects and persists them to SQLite. An event bus connects agents. A lightweight tool registry mediates evidence access. FastAPI provides the dashboard. Configured derived files support replay. Deterministic synthetic states isolate controlled Agent behavior from sensing metrics.

Raw and derived sensing records remain local to the prototype host. When an external OpenAI-compatible model is selected, the Diagnosis Agent sends a minimized structured event context. A local model can replace that endpoint without changing the Agent contract. The same contract carries access control, retention policy, encryption, and explicit resident consent.

\begin{figure}[t]
\centering
\includegraphics[width=\columnwidth]{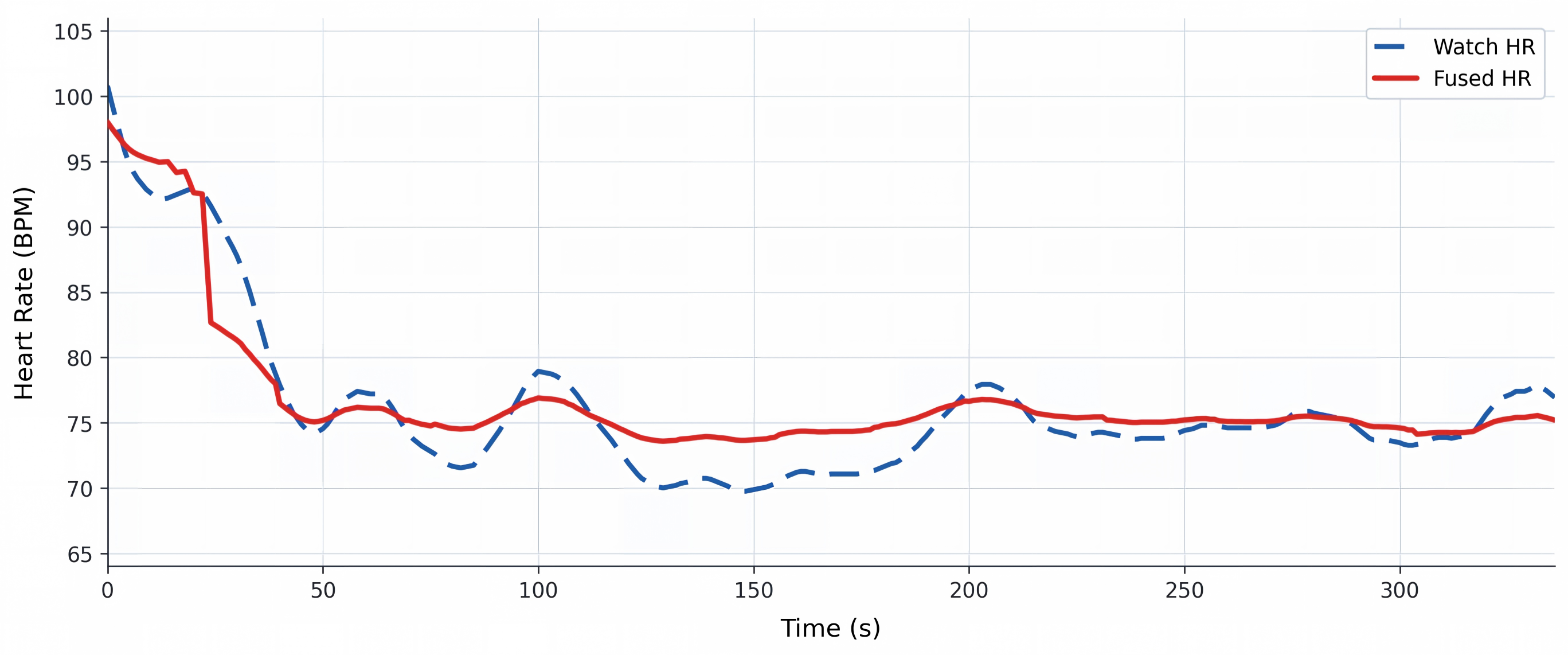}
\caption{Heart-rate tracking example comparing the fused estimate with the watch reference.}
\Description{Line plot showing fused heart-rate tracking compared with a watch heart-rate reference over time.}
\label{fig:heart_tracking}
\vspace{2pt}
\includegraphics[width=\columnwidth]{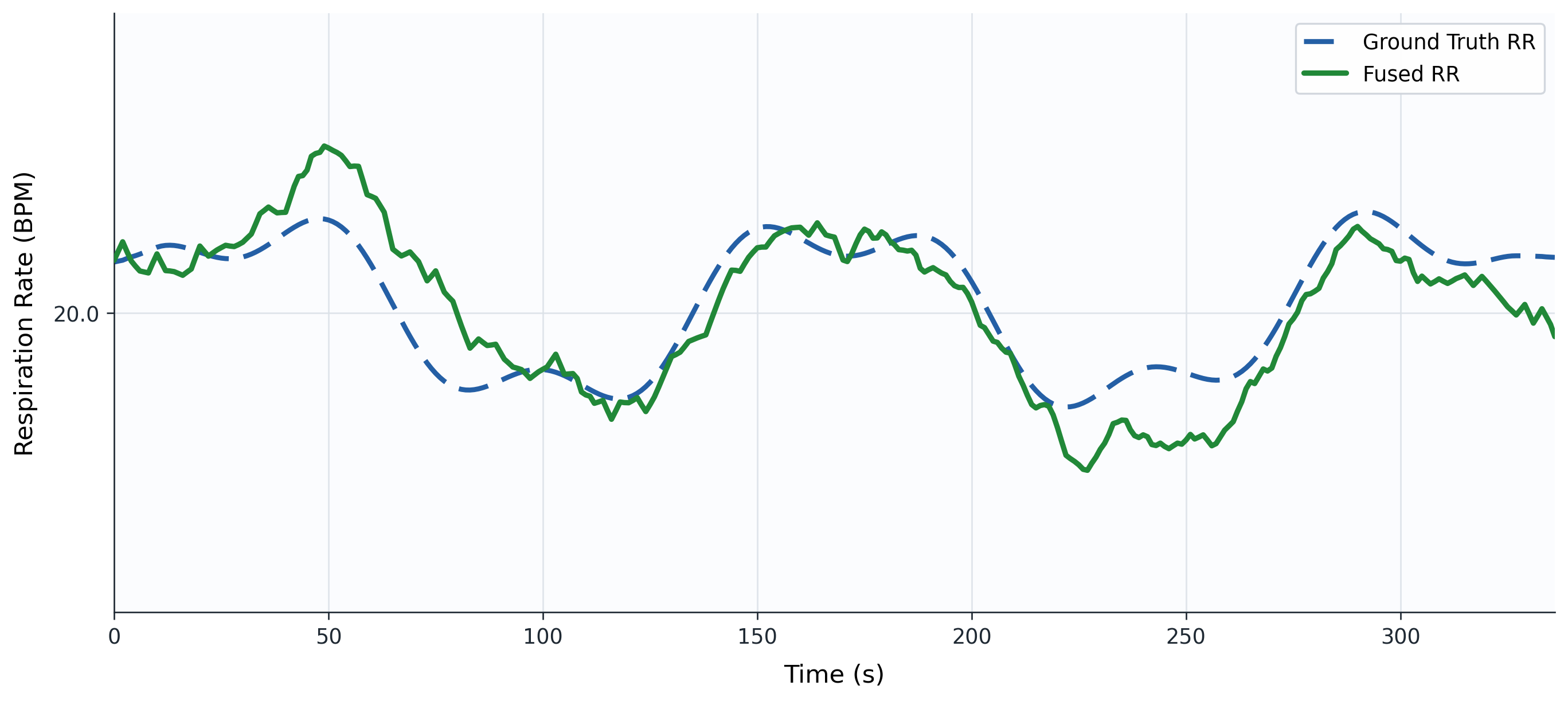}
\caption{Respiratory-rate tracking example comparing the fused estimate with the respiration reference.}
\Description{Line plot showing fused respiratory-rate tracking compared with a reference respiratory rate over time.}
\label{fig:respiration_tracking}
\end{figure}

\addtocounter{table}{1}
\begin{table*}[!t]
\centering
\caption{Disease-relevant indicator coverage of SuperSenseDoctor.}
\label{tab:screening}
\scriptsize
\setlength{\tabcolsep}{3.0pt}
\renewcommand{\arraystretch}{1.05}
\begin{tabularx}{\textwidth}{>{\raggedright\arraybackslash}p{0.225\textwidth}*{6}{>{\centering\arraybackslash}X}}
\toprule
Condition and screening direction &
\shortstack{Respiratory\\rate and pattern} &
\shortstack{Heart\\rate} &
\shortstack{Gait and\\balance} &
\shortstack{Activity and\\mobility} &
\shortstack{Sleep rhythm and\\fragmentation} &
\shortstack{Surface-temp.\\trend} \\
\midrule
COPD exacerbation \cite{chen2025copd}
  & \ScreenYes & \ScreenYes & \ScreenNo & \ScreenYes & \ScreenNo & \ScreenNo \\
Obstructive sleep apnea \cite{kapur2017osa,sequeira2019osahrv}
  & \ScreenYes & \ScreenYes & \ScreenNo & \ScreenNo & \ScreenYes & \ScreenNo \\
Heart-failure decompensation \cite{boehmer2017multisense}
  & \ScreenYes & \ScreenYes & \ScreenNo & \ScreenYes & \ScreenNo & \ScreenNo \\
Parkinson's disease \cite{sun2024digitalpd}
  & \ScreenNo & \ScreenNo & \ScreenYes & \ScreenYes & \ScreenYes & \ScreenNo \\
Cognitive decline and dementia \cite{popp2024passivead,cepukaityte2024dementia}
  & \ScreenNo & \ScreenNo & \ScreenYes & \ScreenYes & \ScreenYes & \ScreenNo \\
Acute respiratory infection \cite{mason2022tempredict}
  & \ScreenYes & \ScreenYes & \ScreenNo & \ScreenYes & \ScreenYes & \ScreenYes \\
\bottomrule
\end{tabularx}
\end{table*}
\addtocounter{table}{-2}

\section{Evaluation}

The evaluation follows the full pipeline. It first checks usable physiological and fall evidence across multiple time intervals. It then tests whether the Agent layer converts standardized situations into records and actions that match policy.

\begin{table}[H]
\centering
\caption{Multimodal sensing validation. Vital-sign rows report MAE and RMSD in bpm.}
\label{tab:sensing}
\footnotesize
\setlength{\tabcolsep}{0pt}
\renewcommand{\arraystretch}{1.02}
\begin{tabularx}{\columnwidth}{@{}>{\centering\arraybackslash}m{0.17\columnwidth}>{\centering\arraybackslash}m{0.21\columnwidth}>{\centering\arraybackslash}Xrr@{}}
\toprule
Task & Ground truth & Method & MAE & RMSD \\
\specialrule{\lightrulewidth}{0pt}{0pt}
\multirow{3}{0.17\columnwidth}{\centering Heart rate} &
\multirow{3}{0.21\columnwidth}{\centering Huawei Watch GT~3} &
WiFi BFI & 4.237 & 4.396 \\
& & mmWave & 4.121 & 4.290 \\
& & Final & \textbf{1.994} & \textbf{3.142} \\
\specialrule{\lightrulewidth}{0pt}{0pt}
\multirow{3}{0.17\columnwidth}{\centering Respiratory rate} &
\multirow{3}{0.21\columnwidth}{\centering Respiration belt} &
WiFi BFI & 2.689 & 3.293 \\
& & mmWave & 1.629 & 1.876 \\
& & Final & \textbf{0.197} & \textbf{0.263} \\
\midrule
Fall recognition & Labeled event replays & BFI and mmWave evidence & \multicolumn{2}{c@{}}{\textbf{\FallRecognitionAccuracyPct\% accuracy}} \\
\bottomrule
\end{tabularx}
\end{table}

\subsection{Multi-Time-Interval Sensing Validation}

The evidence package contains synchronized BFI and mmWave recordings sampled as one-second derived windows. Figures~\ref{fig:heart_tracking} and~\ref{fig:respiration_tracking} show representative fused heart-rate and respiratory-rate traces against their references. We split the larger recording into seven 300~s intervals plus one 249~s interval, and treat the second recording as one 337~s interval. This yields \EvalIntervalCount\ intervals and \EvalStateRows\ state rows. Huawei Watch GT~3 provides the heart-rate reference, while a respiration belt provides the respiratory-rate reference. Fall recognition is evaluated on labeled event replays.

Table~\ref{tab:sensing} summarizes the aggregate sensing metrics. In this offline comparison, the final heart-rate estimate lowers MAE by 51.6\% relative to mmWave and by 52.9\% relative to WiFi BFI. The final respiratory-rate estimate lowers MAE by 87.9\% relative to mmWave and by 92.7\% relative to WiFi BFI. A controlled fall-recognition check reaches \FallRecognitionAccuracyPct\% accuracy as an event-level counterpart. The pipeline carries disagreement-rich evidence into care support while preserving raw modality state for arbitration.

\subsection{Agent Scenario Evaluation}

The Agent benchmark targets the layer where sensing evidence becomes action. It contains a deterministic rule-screening test, a fusion-arbitration test, and a hidden-answer triage test. The deterministic test contains 60 manually specified state objects spanning normal sensing, heart-rate and respiratory deviations, falls, modality conflicts, NLOS, dual-low-confidence input, compound risk, and threshold boundaries. Each case specifies the expected Nurse event set. Fusion cases specify the expected dominant modality or an explicit \texttt{none} verdict.

A separate set of 24 event-centered scenarios tests the virtual triage doctor. Each scenario specifies the expected L0--L4 tier, action channel, evidence anchors, JSON shape, care-support boundary, and severe-risk status. The model receives only standardized event and state evidence. The answer key remains hidden.

The benchmark checks the care loop at the point where sensing becomes action. In the deterministic layer, event screening matches \AgentEventPassed\ of \AgentEventCases\ manually specified cases, and fusion arbitration matches \AgentFusionPassed\ of \AgentFusionCaseCount\ modality-conflict cases. In the model-involved layer, the triage doctor returns the expected tier and action channel in \AgentDiagnosisPassed\ of \AgentDiagnosisScenarioCount\ hidden-answer scenarios, while \AgentTracePassed\ of \AgentDiagnosisScenarioCount\ responses preserve the required structured trace. Overall, using \AgentDiagnosisModelName, the suite passes \AgentChecklistPassed\ of \AgentChecklistItems\ checked criteria, reaching \AgentChecklistPassPct\%. The deterministic checks run with a \AgentDetPNinetyFiveMs~ms P95 latency, and the model-involved triage path runs with a \AgentDiagnosisPNinetyFiveMs~ms P95 latency. The Diagnosis Agent therefore acts only after the deterministic gate creates an event, then returns a typed decision with cited evidence and an action channel.

\section{Future Outlook and Development}

The current prototype turns multimodal room sensing into an inspectable care-support path. Configured BFI and mmWave observations become state evidence before conflicts, NLOS conditions, low-confidence states, or falls reach the Agent layer. Viewers can trace the Nurse event, evidence, action tier, delivery channel, and persisted \texttt{EpisodeLog}. Privacy and human oversight remain defaults: raw streams stay local, external models receive compact structured events, and high-risk actions require authorized confirmation.

The next step is to let this state grow with the resident. Personalized baselines can connect respiratory, cardiac, activity, sleep, and surface-temperature patterns across days and rooms. The Nurse Agent can then detect change relative to routine, while the Diagnosis Agent retrieves only the episode history needed for interpretation. Clinician review can refine baseline ranges, evidence priority, and follow-up policy.

This creates a path from single-event response toward longitudinal screening support. Table~\ref{tab:screening} maps respiratory, cardiac, gait, activity, sleep, and surface-temperature cues to condition families such as COPD exacerbation, sleep apnea, heart-failure burden, Parkinsonian movement change, cognitive decline, and acute respiratory infection. New screening paths can reuse the same multimodal state, event gate, Agent tools, and episode log while leaving clinical interpretation to qualified professionals. These results move \textbf{SuperSenseDoctor} toward a home that observes continuously, reasons selectively, and routes timely evidence to people who can act.

\begingroup
\renewcommand{\bibfont}{\scriptsize}
\setlength{\bibsep}{0pt}
\clearpage
\bibliographystyle{ACM-Reference-Format}
\bibliography{SuperSenseDoctor_references}
\endgroup

\end{document}